\documentclass[referee]{mn2e}
\usepackage{graphicx}
\newcommand{\etal}{{\rm et al.~}}           
\newcommand{\NH}{\mbox{${\rm N_H}$ }}       
\newcommand{\NHunits}{\mbox{$~{\rm cm}^{-2}$} }

\begin{document}

\title {Soft X-ray properties of a high redshift
sample of QSOs observed with {\it ROSAT}}
\author[Dewangan et al.]
{G. C. Dewangan,$^1$  K. P. Singh,$^1$, K. F. Gunn,$^2$, A. M. Newsam$^3$, 
\newauthor I. M. McHardy,$^2$, and L. R. Jones,$^4$  \\
$~^1$ Department of Astronomy \& Astrophysics, Tata Institute of Fundamental Research, Mumbai, India 400~005 \\
$^2$Department of Physics \& Astronomy, University of Southampton, Highfield, Southampton, SO17 1BJ \\
$~3$Astrophysics Research Institute, Liverpool John Moores University, Liverpool, CH41 1LD, UK \\
$~^4$School of Physics \& Astronomy, University of Birmingham, Birmingham B15~2TT, U.K. \\ }

\maketitle
\label{firstpage}
\begin{abstract}
In order to study systematically the soft X-ray emission 
of Active Galactic Nuclei (AGNs) at medium to high redshifts, 
we have analyzed {\it ROSAT} PSPC and HRI data of QSOs at $0.26\le z \le3.43$
selected from the second deepest {\it ROSAT} PSPC survey carried out in 1991-1993 by 
McHardy \etal (1998). Our sample of 22 type~1 QSOs is nearly complete above a flux limit of $1.4\times10^{-14}{\rm~erg~cm^{-2}~s^{-1}}$ in the $0.1-2{\rm~keV}$ band.
Of these,
nine QSOs show long term ($\sim 2{\rm~yr}$) X-ray variability by 
a factor of $1.5-3.5$. 
Significant excess absorption above the 
Galactic column is seen in three QSOs.  
The soft X-ray photon index of the QSOs 
ranges from $1.4$ to $3.7$. Three QSOs have steep soft X-ray spectra ($\Gamma_{X} > 3.0$),
one of which is a narrow-line QSO -- a high
luminosity version of narrow-line Seyfert~1 galaxies. The average photon index ($<\Gamma_{X}>$) is 
$2.40\pm0.09 $ (with a dispersion of $0.57$) in the $0.1-2{\rm~keV}$ band. 
The average QSO spectra in four redshift bins flatten from an average photon index of 
$<\Gamma_{X}> \sim 2.53$ at $0.25 \le z \le 1$ to $<\Gamma_{X}> \sim 2$ at $2 \le z \le 3.4$.
The flattening 
of the average photon 
index  can 
be understood in terms of the redshift effect of the mean intrinsic QSO 
spectrum consisting of two components -- a soft X-ray excess 
and a power-law component.  
We have also studied optical spectra of 12 of the 22 QSOs.
\end{abstract}
\begin{keywords}
galaxies:active -- galaxies:nuclei -- X-rays: galaxies -- X-rays: sources 
\end{keywords}

\section{Introduction}
QSOs or quasars represent the high luminosity end of a class of
 objects known as active galactic nuclei (AGNs). They are the most 
luminous continuously emitting objects in the universe and emit over the entire range of
electromagnetic waves. It is widely believed that high energy 
emission (X-rays and $\gamma$-rays) from an AGN originates 
in the innermost region of accretion disk around 
a super-massive black hole (SMBH) and much of the low energy 
emission e.g., infrared, optical, ultraviolet, 
is due to the reprocessing of high energy photons in a medium 
surrounding the accretion disk. While the atomic 
line emission in the optical, UV, and X-ray probe 
the circumnuclear medium surrounding the SMBH, the shape 
of the X-ray spectrum described by photon index ($\Gamma_{X}$) 
is a critical parameter to constrain competing models for 
the mechanisms of X-ray continuum emission. 

The increased sensitivity of the position sensitive 
proportional counter (PSPC) on-board {\it ROSAT} compared 
to earlier missions allowed a significant improvement 
in the study of the soft X-ray emission of AGNs. 
The X-ray spectral shape of quasars has been studied extensively using earlier
missions such as {\it HEAO-I, Einstein, EXOSAT, and Ginga} (e.g., Mushotzky 1984; Wilkes \& Elvis 1987; Canizares \& White 1989; Comastri \etal 1992; Lawson \etal 1992; Williams \etal 1992). Some of these missions were not sensitive below $2{\rm~keV}$. These earlier studies suggested that the X-ray emission 
of quasars is well described by a power-law with photon index $\Gamma_X \sim 1.5$ for radio-loud quasars and $\sim 2.0$ for radio quiet quasars.
Large samples of AGNs have been studied by Walter \& Fink (1993), 
Wang, Brinkmann \& Bergeron (1996), and by Rush \etal (1996) 
using the {\it ROSAT} PSPC. However, most of the 
objects studied in the above papers are nearby and intrinsically
bright AGNs. Also the AGN samples studied in the above papers 
are not complete and the derived results may be 
biased by selection effects. Laor \etal (1994, 1997) studied X-ray spectra of a 
complete sample of 23 quasars from the Bright Quasar Survey (BQS) 
with $z\le0.4$ and ${\rm \NH} < 1.9\times10^{20}{\rm~cm^{-2}}$. 
They found an average photon index of $2.63\pm0.07$ for their 
complete sample. Reeves \& Turner (2000) studied a larger 
sample of 62 quasars  with redshift in the range 0.06 -- 4.3 
and ${\rm M_{V}<-23}$. They found an average photon index, 
$\Gamma_{X} = 1.66\pm0.04$ for 35 radio-loud quasars and 
$\Gamma_{X} = 1.89\pm0.05$ for 27 radio-quiet quasars in 
their sample. Although the sample of Laor \etal (1997) is 
complete, it is confined to the nearby universe. 
The  sample of Reeves \& Turner (2000) 
includes some high redshift quasars but most of the quasars 
are nearby and the sample is not complete. 
 
We present the results of a detailed spectral and timing 
analysis of a nearly complete sample of 22 type~1 QSOs obtained from 
the {\it ROSAT} PSPC deep survey of McHardy \etal (1998). The basic parameters of the QSOs are listed in Table~\ref{basic_par}. We will
refer to all the objects as QSO,
independent of their luminosity, but note that some have low X-ray luminosity (see Table~\ref{fit91_result}).
The aims are to extend the study of soft X-ray properties 
of QSOs at higher redshifts and to investigate whether QSOs at high redshifts have 
excess soft X-ray emission similar to that seen in the narrow-line Seyfert~1 galaxies.  
The outline of the paper is as follows. In $\S 2$ we 
describe the sample; in $\S 3$ we describe the X-ray observations and analysis. 
Section 4 deals with optical spectroscopy. 
In $\S 5$, we compare our results with 
other studies and discuss some of the implications. Finally, 
we conclude our study in $\S 6$.

Throughout the paper, luminosities are calculated assuming  
isotropic emission, a Hubble
constant of $H_{0}=75{\rm~km~s^{-1}~Mpc^{-1}}$ and a deceleration
parameter of $q_{0}=0$ unless otherwise specified.

{\begin{figure*}
	\centering
	\includegraphics[angle=0,width=15cm]{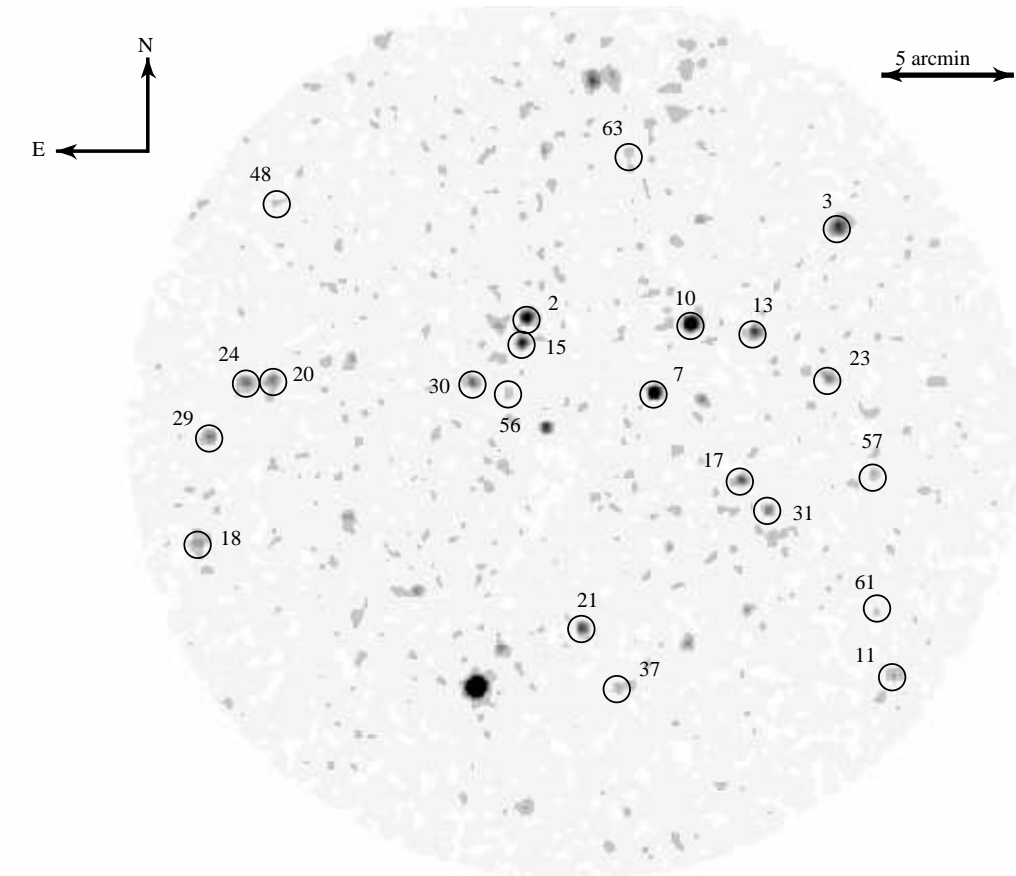}
	\caption{MJM QSOs observed with the {\it ROSAT} HRI. The image was binned by $16{\rm~pixels}$ ($8{\rm~arcsec}$) and then smoothed by convolving with a Gaussian of $\sigma = 2{\rm~pixels}$.}
	\label{hri_qso}
\end{figure*}}

\begin{table*}
\caption{Basic parameters of QSOs}
\label{basic_par}
\begin{flushleft}
\begin{tabular}{@{}ccccccccc}
\hline
MJM$^1$ No. & {\it ROSAT} name & \multicolumn{2}{c}{HRI position$^2$} & \multicolumn{2}{c}{Optical position$^3$} & z$^3$ & $m_{R}^3$ & Radio$^4$  \\
            &           & $\alpha (2000)$ & $\delta (2000)$ & $\alpha (2000)$ & $\delta (2000)$      &               & \\ \hline
2 &  RX J1334.7+3800 & 13 34 41.87 & 38 00 11.4 &  13 34 41.82 &    38 00 11.3 &   0.26 & 18.69   & (r)       \\
3 &  RX J1333.7+3803 & 13 33 42.41 & 38 03 35.3 &  13 33 42.36 &    38 03 36.3 &   1.069 & 18.60  &  \\
7 &  RX J1334.3+3757 &  13 34 17.52 & 37 57 22.1 &  13 34 17.52 &    37 57 22.4 &   1.14 &18.35    &  \\
10 &  RX J1334.2+3759 & 13 34 10.57 & 37 59 56.4  & 13 34 10.62 &    37 59 56.3 &   0.38 &19.55   &   \\
11 &  RX J1333.5+3746 & 13 33 32.32 & 37 46 42.4  & 13 33 32.01 &    37 46 41.1 &   0.826 &20.74  &   \\
13 &  RX J1333.9+3759 & 13 33 58.49 & 37 59 39.3  & 13 33 58.55 &    37 59 38.2 &   1.61 &21.10   &   \\
15 &  RX J1334.7+3759 & 13 34 42.81 & 37 59 15.0  & 13 34 42.77 &    37 59 15.0 &   1.14 &19.83   &   \\
17 &  RX J1334.1+3754 & 13 34 00.96 & 37 54 04.7  & 13 34 01.03 &    37 54 04.9 &   1.64 &21.12   &   \\
18 &  RX J1335.7+3751 & 13 35 44.46 & 37 51 43.3  & 13 35 44.66 &    37 51 40.8 &   1.621 &20.25  &   \\
20 &  RX J1335.5+3757 & 13 35 30.26 & 37 57 50.5  & 13 35 30.30 &    37 57 50.0 &   1.39 &20.86   &   \\
21 &  RX J1334.5+3748 & 13 34 31.30 & 37 48 31.2  & 13 34 31.33 &    37 48 31.4 &   1.359 &22.27 &    \\
23 &  RX J1333.7+3757 & 13 33 44.34 & 37 57 53.6  & 13 33 44.27 &    37 57 52.6 &   0.97 & 20.37 &   \\
24 &  RX J1335.6+3757 & 13 35:35.48 & 37:57:46.3  & 13 35 35.48 &    37 57 46.2 &   1.63 &20.21  &  \\
29 &  RX J1335.7+3755 & 13 35 42.38 & 37 55 43.6  & 13 35 42.51 &    37 55 41.8 &   1.90 &19.30   &  \\
30 &  RX J1334.9+3757 & 13 34 52.23 & 37 57 44.7  & 13 34 52.16 &    37 57 44.8 &   1.89 &20.47   & (r)\\
31 &  RX J1333.9+3752 & 13 33 55.94 & 37 52 57.5  & 13 33 55.81 &    37 52 58.6 &  2.14 &20.32  &  \\
37 &  RX J1334.4+3746 & 13 34 24.43 & 37 46 15.9  & 13 34 24.57 &    37 46 15.2 &   1.570 &20.11 &  \\
48 &  RX J1335.5+3804 & 13 35 30.00 & 38 04 31.5  & 13 35 29.69 &    38 04 32.7 &  0.692 &19.17  &  \\
56 &  RX J1334.7+3757 & 13 34 45.43 & 37 57 21.5  & 13 34 45.35 &    37 57 22.8 &   1.89 &20.36  &  \\
57 &  RX J1333.6+3754 & 13 33 35.69 & 37 54 16.7  & 13 33 35.62 &    37 54 13.2 &   1.525 &21.68 &  \\
61 &  RX J1333.6+3749 & --  & -- & 13 33 34.86 &    +37 49 16.92 &   3.43 &19.73  &  \\
63 &  RX J1334.4+3806 & 13:34:22.44 &  38:06:21.5  & 13 34 22.24 &    38 06 20.1 &   2.593 &21.40 &  \\
\hline 
\end{tabular}
\newline
$~^1$Identification number in the catalog of McHardy \etal (1998). \\
$~^2$Derived from {\it ROSAT} HRI observation of 1997. \\
$~^3$McHardy \etal (1998) \\
$~^4$ (r) indicates that the source is detected in the preliminary 
20-cm radio map. \\
\end{flushleft}
\end{table*}

\section{The QSO sample}
\begin{figure*}
	\centering
	\includegraphics[angle=0,width=10cm]{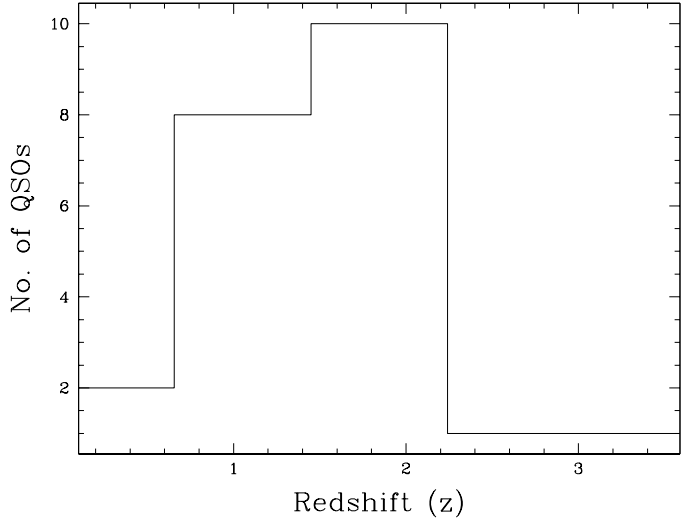}
	\label{histo_z}
	\caption{Redshift distribution of the 22 MJM QSOs}
\end{figure*}

The QSOs were selected from the second deepest {\it ROSAT} 
PSPC survey of McHardy \etal (1998) which covers a 
circular region of radius $15{\rm~arcmin}$ of the sky towards the direction 
 where the Galactic \NH is low 
($\sim 7.9\times10^{19}{\rm~cm^{-2}}$). The {\it ROSAT} 
survey consists of two observations with 
PSPC and one observation with HRI (see below). There are 
32 QSOs identified to a flux limit of 
$\sim2\times10^{-15}{\rm~erg~cm^{-2}~s^{-1}}$ in the 
energy band of 0.5--2.0~keV (see Jones \etal 1997; 
McHardy \etal 1998). In order to be able to derive the shape 
of the soft X-ray spectrum of the individual QSOs from 
the single observation with the longest exposure and thus to avoid any effect due to the variation of soft X-ray 
emission over long time scales, we have selected 22 
QSOs and excluded the faintest 10 QSOs which are not
detected above $5\sigma$ level in any of the two PSPC 
observations. The soft X-ray selected sample of
22 QSOs (type~1) is nearly complete to a flux limit of 
$\sim1.4\times10^{-14}{\rm~erg~cm^{-2}~s^{-1}}$ in the 
energy band of 0.1$-$2.0~keV (corresponding to $\sim 5\times 10^{-15}{\rm~erg~cm^{-2}~s^{-1}}$ in the $0.5-2{\rm~keV}$ energy band). There are two X-ray sources, MJM~9 and 
MJM~54 brighter than the above flux limit but the optical nature of these objects is not known (McHardy \etal 1998) due to either lack of an optical counterpart or ambiguity in the identification.  
The {\it ROSAT} HRI image of the 
circular region of radius ${\rm 15~arcmin}$ is shown 
in Figure~\ref{hri_qso}. The HRI image was binned by $16{\rm~pixels}$ ($8{\rm~arcsec}$) and then smoothed by convolving with a Gaussian of $\sigma = 2{\rm~pixels}$. All the 22 QSOs are marked in Fig.~\ref{hri_qso}. The 
centres of the small circles are the optical positions of the 
QSOs as identified by McHardy \etal (1998). X-ray sources 
shown in Fig.~\ref{hri_qso}, were classified as QSOs based on their 
optical spectra. All the 22 QSOs show broad (full width at 
half maximum (FWHM) $>1000{\rm~km~s^{-1}}$) permitted 
emission lines (Jones \etal 1997; McHardy \etal 1998). The hardness ratio ($HR$), 
defined as $HR = \frac{H-S}{H+S}$, where H and S are the count rates in the 0.5--2.0~keV 
and 0.1--0.5~keV bands, of the QSOs 
spans a range of $\sim0.14-1.78$. In Figure~\ref{histo_z}, we show 
the distribution of the QSOs in the sample as a function 
of redshift and R mag. 
The number of QSOs increases with magnitude up to $R=20.5{\rm~mag}$ 
and falls at fainter magnitudes (see Fig.~\ref{histo_z}). The QSO sample 
is optically complete at $R<21$ but it is likely that a 
few fainter QSOs have not been detected (McHardy \etal 1998). 
The redshift of the QSOs ranges
from $0.26$ to $3.43$. 

\section{X-ray observations and analysis}
\subsection{X-ray Observations}
The region of the sky containing the QSOs was observed twice with
the {\it ROSAT} (Tr\"{u}mper 1983) Position Sensitive Proportional
Counter (PSPC) during 1991--1993, and once with the High Resolution Imager (HRI)
(Pfeffermann et al. 1987) in 1997 June 19--July 16. The two PSPC observations carried
out during 1991 June 23--26, and 1993 June--July together comprise the second
deepest {\it ROSAT} PSPC survey. These observations were targeted at
$\alpha(2000)=13^{h}34^{m}37.0^{s}$, $\delta(2000)=+37{\degr}54{\arcmin}44{\arcsec}$
in the sky with an extremely low obscuration due to matter in our
Galaxy ($\NH=7.9\times10^{19}\NHunits$).
The PSPC deep survey data has already been reported by McHardy \etal (1998).
The HRI observation was also a deep survey (exposure time = 201513~s) and was
carried out in the same region
of sky as the PSPC deep surveys.
The details of the {\it ROSAT}
observations are given in Table~\ref{rosat_obs}.  All the QSOs lie within $15\arcmin$ from the
field centre.
\begin{table*}
\caption{Details of the {\it ROSAT} observations of the QSOs.}
\label{rosat_obs}
\begin{flushleft}
\begin{tabular}{lllllll}
\hline
Serial & Sequence & Instrument &  Start Time & End Time & Exposure   \\
No. & No. &   & Y, M, D, UT & Y, M, D, UT & Time (s)  \\
\hline
1. & RH900717N00& HRI &  1997 06 04  16:12:58 &1997 07 13 22:26:43 & 201513  \\
2. &RP900626N00 &PSPC   &1993 06 19 22:24:46&1993 07 16 23:06:28 &37658 &  \\
3. & RP700283N00 &PSPC   & 1991 06 23 20:59:40 &1991 06 26 20:44:16  &71803  \\
\hline
\end{tabular}
\end{flushleft}
\end{table*}

\subsection{Optical Identification}
\begin{figure*}
        \centering
	\vspace*{-1.2cm}
        \includegraphics[width=\textwidth, bb = 30 30 575 575]{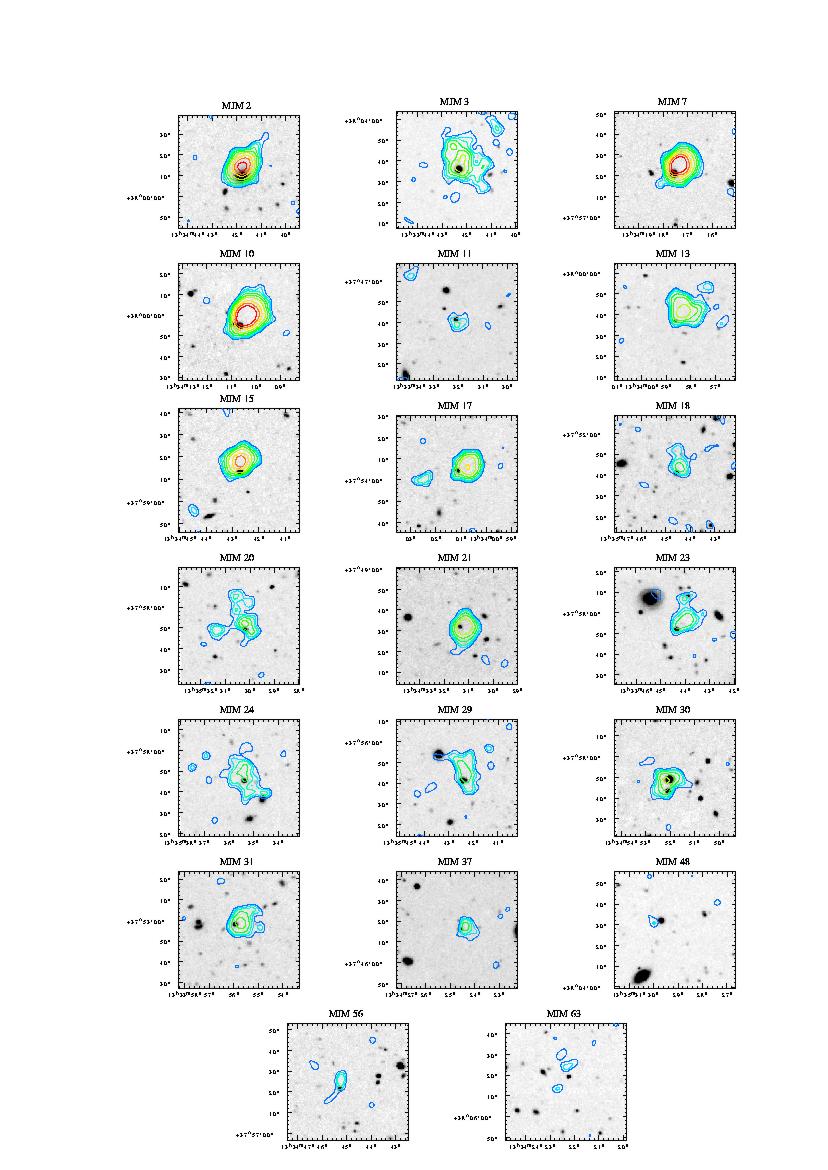}
        \label{hri_overlay}
        \caption{Optical R band images of the QSOs overlaid by {\it ROSAT} HRI contours. In each optical image the central object is the QSO. The X-ray contours are drawn at intensity levels $0.119$, $0.137$, $0.155$, $0.191$, $0.245$, $0.335$, $0.515$, $0.695$, and $0.965{\rm~count~pixel^{-1}}$}
\end{figure*}

The X-ray sources, presented here, were optically identified 
with QSOs using {\it ROSAT} PSPC and optical images, and optical spectra by 
McHardy \etal (1998). Here we have confirmed their identification using 
high resolution HRI X-ray images(central full width half maximum of $\sim4\arcsec$).
The analysis of HRI X-ray images has been carried out using the
PROS\footnote[1]{The PROS software package provided by the {\it ROSAT}  Science
Data Center at Smithsonian Astrophysical Observatory.} software package.
X-ray images were extracted from the observations listed in Table~\ref{rosat_obs} and
were smoothed by convolving with a Gaussian of
$\sigma$=$2\arcsec$. We have overlaid the contours of the
smoothed HRI image of each QSO onto the corresponding optical image. 
The optical R band images  the QSOs were obtained
from the 2.5-m Isaac Newton Telescope (INT) using the wide-field camera on the
night of 1999 April 19.
The pixel scale is 0.333~arcsec per pixel and the
seeing (FWHM) was $\sim1.3\arcsec$.
The overlays of X-ray contours onto the R-band images are shown in Figure~\ref{hri_overlay}. The QSO MJM~61 was not detected in the HRI 
observation, therefore it is not shown in Fig.~\ref{hri_overlay}. 
The overlays reveal that X-ray emission from 
some of the QSOs is contaminated with that due to neighbouring sources.
In particular, the X-ray emission from the QSOs MJM~18, 20, 23, and 24 may 
be contaminated significantly, however, the dominant X-ray emission is that 
due to QSOs. In the following, we have attributed the X-ray emission seen with
the PSPC to the
individual QSOs due to the difficulty in separating 
out the small contribution of neighbouring sources in the lower resolution 
(FWHM $\sim 25{\rm~arcsec}$) PSPC images.

The HEASOFT v5.0.1 package was used to analyze the PSPC X-ray data 
of all QSOs.
We have estimated the X-ray count rates for all the QSOs in our sample 
from the PSPC observations of 1991 and 1993. The total source counts 
for each QSO were obtained in the energy band of 0.1--2.0 keV
from the unsmoothed PSPC images using a circular region 
centered on the peak positions, and after subtracting the background
estimated from 3-4 nearby circular regions. Note that for a few 
objects e.g., MJM~2, MJM~15 the centres of the circular region were 
shifted by a few arcsec in order to avoid the contamination from the 
neighboring objects. The radii of circles, used to extract the 
source counts, vary from $45-90{\rm~arcsec}$. This was necessitated due 
to the different off-axis position of individual QSOs and to avoid 
contamination with the nearby X-ray sources. The background counts were
estimated separately for individual objects using 3-4 nearby circular
regions of radii $1.5-2.5{\rm~arcmin}$ in order to avoid any spatial
variation of the background intensity. Net source counts were derived
by subtracting appropriately scaled background counts. The count rates 
of all the QSOs, derived from the observations of 1991 and 1993, are listed 
in Table~\ref{cr91_93}. Also listed in Table~\ref{cr91_93} are the hardness ratio (HR) values
where HR is defined as 
the ratio of the source counts in the energy band of 0.5--2.0~keV to 
that in the energy band 0.1--0.5~keV. The QSOs MJM~31, 56, and 63,  
are not detected during the observation of 1993. 

\begin{table}
\caption{Observed count rates and hardness ratios of MJM QSOs}
\label{cr91_93}
\begin{flushleft}
\begin{tabular}{@{}lccccc}
\hline
MJM No.$^1$ & \multicolumn{2}{c}{1991} &  \multicolumn{2}{c}{1993} & Variability \\
        & count rate & HR   & count rate & HR & factor \\
        & $10^{-3}{\rm~cnt~s^{-1}}$ &    &$10^{-3}{\rm~cnt~s^{-1}}$ & & \\ 
\hline
2       & $9.52\pm0.54$  & $-0.213\pm$ & $18.4\pm0.88$  & $ -0.127\pm0.046$ & $\sim 2$ \\
3($*$)  & $16.2\pm0.70$  & $-0.564\pm0.038$ & $9.92\pm1.05$  & $-0.531\pm0.085$ & -- \\
7       & $7.25\pm0.46$  & $-0.523\pm0.052$ & $14.49\pm0.84$ & $-0.647\pm0.049$ & $\sim 3$ \\
10      & $19.16\pm0.66$ & $-0.768\pm0.025$ & $16.91\pm0.85$ & $-0.794\pm0.040$ & -- \\
11($*$) & $5.60\pm0.45$  & $-0.216\pm0.077$ & $5.31\pm0.64$  & $-0.354\pm0.114$ & -- \\
13($*$) & $5.41\pm0.43$  & $-0.412\pm0.068$ & $7.70\pm0.93$  & $-0.269\pm0.108$ & -- \\
15      & $11.12\pm0.55$ & $-0.627\pm0.039$ & $5.06\pm0.66$  & $-0.621\pm0.127$ & $\sim 2$\\
17($*$) & $3.27\pm0.41$  & $-0.244\pm0.114$ & $6.08\pm0.73$  & $-0.247\pm0.111$ & $\sim 2$ \\
18      & $3.18\pm0.46$  & $-0.321\pm0.128$ & $3.88\pm0.75$  & $-0.083\pm0.187$ & --\\
20($*$) & $3.95\pm0.41$  & $-0.485\pm0.087$ & $2.78\pm0.56$  & $-0.221\pm0.195$ & -- \\
21      & $1.66\pm0.37$  & $0.109\pm0.216$ & $3.16\pm0.61$  & $0.144\pm0.204$ & $\sim 1.5$ \\
23($*$) & $9.25\pm0.63$  & $-0.675\pm0.054$ & $6.50\pm0.77$  & $-0.600\pm0.103$ & -- \\
24($*$) & $3.56\pm0.41$  & $-0.331\pm0.127$ & $4.35\pm0.59$  & $-0.264\pm0.132$ & -- \\
29($*$) & $6.97\pm0.63$  & $-0.632\pm0.085$ & $6.94\pm0.74$  & $-0.602\pm0.095$ & -- \\
30($*$) & $1.80\pm0.36$  & $0.047\pm0.156$ & $3.67\pm0.58$  & $-0.269\pm0.154$ & $\sim 1.8$ \\
31      & $1.58\pm0.34$  & $0.251\pm0.221$ & $<1.7$         & --              & -- \\
37      & $3.01\pm0.45$  & $-0.687\pm0.125$ & $2.83\pm0.76$  & $0.638\pm0.244$ & -- \\
48      & $4.95\pm0.51$  & $-0.532\pm0.095$ & $2.54\pm0.83$  & $-0.021\pm0.325$ & $\sim 2$\\
56      & $2.83\pm0.41$  & $-0.600\pm0.120$ & $<0.65$ & --   & $$ \\
57($*$) & $4.03\pm0.47$  & $-0.617\pm0.108$ & $3.49\pm0.77$  & $-0.305\pm0.203$ & -- \\
61      & $4.23\pm0.58$  & $-0.372\pm0.135$ & $3.16\pm0.75$  & $-0.381\pm0.213$ & --\\
63($*$) & $2.86\pm0.40$  & $-0.436\pm0.146$ & $<1.56$ & --   & $$ \\
 \hline
\end{tabular} 
\newline
$^1$($*$) indicates that the X-ray emission form the source
is contaminated by that due to neighbouring sources. The 
contamination could be as high as $\sim50\%$ in case of
MJM~20, 30, 55, 63, and 75. In other sources, the contamination
is $\sim10\%$ or less. \\
\end{flushleft}
\end{table}

\subsection{X-ray Light Curves}
In order to investigate the time variability of  soft X-ray emission from
the QSOs, we have extracted light curves from the {\it ROSAT} PSPC
observations. The light curves for the source and the background were extracted
using the `xselect' package with time bins of 2000~s and in the PSPC energy
band of 0.1--2.0 keV containing all the X-ray photons falling within 
``good time intervals".  The source regions
and the background regions were the same as described above.  
It was found that the background 
was variable by a factor of $\sim2-3$ during the observation of 
1993 June 19. Therefore, the light 
curves obtained from the 1993 observation are not suitable for 
variability studies. During the observation of 1991 June 23, the background 
was reasonably constant. 
The background
subtractions were carried out after appropriately scaling the background
light curve to have the same area as the source extraction area. 
In order to test the constancy 
of the background light curves and soft X-ray variability of the 
QSOs, we have fitted constant count rates to the background light 
curves and source light curves. 
From the best-fit minimum $\chi^2$ values, it was found  
that background count rates 
obtained separately for each QSO are reasonably constant.
It was also found that all QSOs, except MJM~10, do not show 
significant short-term variability during the observation of 1993 June-July. 
The soft X-ray variability characteristics of the QSO MJM 10 has been 
discussed by Dewangan \etal (2001). 

   A comparison of count rates during the observations 1991 
   and 1993 (see Table~\ref{cr91_93}) reveals that 9 out of 22 QSOs show 
   long term (time scale $\sim 2{\rm~yr}$) variability. The QSOs
   MJM~2, 15, 17, and 48 varied by a factor
   of 2 within 2~yr while the QSOs MJM~7, 21, and, 30 varied by 
   factor of $\sim3$, $\sim 1.5$, and $\sim1.8$, respectively.
   The two QSOs MJM~56, and MJM~63, which are not detected above 
   $3\sigma$ level during the observation of 1993, varied by 
   factor of more than 3.6 and 1.3, respectively.

\subsection{X-ray Spectral Analysis}
Photon energy spectra of all the QSOs in our sample  were accumulated 
from their PSPC observation of 1991.  
The same regions for the source and 
the background,  
as described above ($\S 4.1$), were used. To improve
the statistics, the {\it ROSAT} PSPC pulse height 
data obtained in 256 pulse height channels for each QSO
were appropriately re-grouped to have at least 20 counts in each channel.
As examples, X-ray spectra of two QSOs  
thus obtained from the observation of 1991 are shown in 
Figure~\ref{xray_spectra}\footnote{Rest of the X-ray spectra can be made available  on request to G. C. Dewangan ({\tt email: gulab@tifr.res.in})}. 

We used the XSPEC (Version 11.0) spectral analysis package to
fit the data with spectral models.
An appropriate response matrix
 and an auxiliary response file of the
effective area of the telescope were used to define the energy response of
the PSPC. The {\it ROSAT} PSPC spectra of QSOs shown in Fig.~\ref{xray_spectra}, were 
used for fitting
spectral models.  The X-ray spectrum of each QSO
 in the sample was first fitted
with a redshifted power-law model with photon index, $\Gamma_{X}$,
and absorption due to an intervening medium (Model A) with the absorption cross-sections as
given by Balucinska-Church \& McCammon (1992) and using the method of
$\chi^{2}$-minimization. The results of this fitting and the best-fit
spectral model parameters are shown in Table~\ref{fit91_result}. The errors quoted 
were calculated at the $90\%$ confidence level based on  $\chi^2_{\rm min }$+2.71. 
This simple model is a good fit to all the individual QSO spectra
as evidenced by the minimum reduced $\chi^{2}$  
values. The absorbing column densities derived by fitting the power-law model to the spectra of MJM~2, MJM~3, MJM~11, MJM~13, MJM~17, MJM~18, MJM~20, MJM~21, MJM~23, MJM~24, MJM~31, MJM~37, MJM~56, MJM~57, and MJM~61 are similar within errors to that due to our own 
Galaxy measured from 21-cm radio observations along the 
line of sight to the source (Dickey \& Lockman 1990), however, the error bars in the derived ${\rm N_{H}}$ are quite large for some objects. To better constrain the photon indices for these QSOs, we next fixed the absorbing column density to the Galactic value and fitted the power-law model (Model B) to each QSO. The results of this fitting are also given in Table~\ref{fit91_result}. 

The absorbing column densities, 
$20.7_{-9.7}^{+13.6}\times 10^{19}{\rm~cm^{-2}}$, $18.1_{-4.7}^{+6.0}\times 10^{19}{\rm~cm^{-2}}$, $18.7_{-7.4}^{+9.8}\times 10^{19}{\rm~cm^{-2}}$, derived 
for the QSOs MJM~7, MJM~10, and MJM~15 by fitting absorbed power-law model are in excess of 
the Galactic value 
($7.9\times10^{19}{\rm cm^{-2}}$)  measured along their line of sight. 
This indicates that all the X-ray absorption
may not be only due to matter in our own Galaxy but also due to matter
local to the source. 
In order to
verify the presence of excess absorption, we fixed the neutral hydrogen
column density to the Galactic value and fitted the absorbed power-law
 model (Model B). In each case, although   
the fit is acceptable, there is an increase in the value of $\chi^{2}$ ($\Delta \chi^2 = 5.33$ 
for 23 dof in case of MJM~7; $\Delta \chi^2 = 16.46$ for 25 dof in case of MJM~10; 
$\Delta \chi^2 = 6.648$ for 15 dof in case of MJM~15). 
To estimate the excess absorption local to each of the three QSOs, we introduced an
additional component for absorption local to the source in our model B (Model C -- redshifted power-law modified with Galactic as well as intrinsic absorption local to the source) and fitted to PSPC spectrum of the each of the three QSOs. The derived excess absorption, $\Delta N_H$, are 
$9.7_{-8.1}^{+12.0}\times10^{20}{\rm ~cm^{-2}}$ 
for MJM~7, $2.8_{-0.8}^{+1.7}\times10^{20}{\rm ~cm^{-2}}$ for MJM~10, and 
$9.5_{-7.4}^{+10.0}\times10^{20}{\rm ~cm^{-2}}$ for MJM~15. The best-fit 
photon indices ($\Gamma_{X}$) are $3.1_{-0.6}^{+0.8}$, $3.7_{-0.3}^{+0.3}$, 
and $3.3_{-0.5}^{+0.7}$ for MJM~7, MJM~10, and MJM~15, respectively.  
We measure the statistical significance of the reduction in the best-fit $\chi^{2}$ with 
addition of an intrinsic absorption component, described by \NH as a free parameter, 
using the F-test (Bevington 1969). The calculated values of F-statistic values and the 
corresponding probabilities are given in Table~\ref{ftest} for the QSOs MJM 7, 10, and 15. 
The F-statistics probability is the probability that the reduction in $\chi^2$ is not 
statistically significant. As can be seen the Table~\ref{ftest}, at $95\%$ level, the presence of 
additional absorption component local to the source is significant for the QSOs MJM~7,  MJM~10 and MJM~15.  

\begin{table*}
\caption[]{Spectral model parameters of QSOs derived from the observation 1991}
\label{fit91_result}
\begin{flushleft}
\begin{tabular}{cccccccc}
\hline
MJM No. & Model$^a$ & ${\rm N_{H}}$ & $\Gamma_{X}$ & $f_{obs}^b$ & $L_{unabs}^c$  & Minimum  & dof \\
        &       & $10^{19}{\rm~cm^{-2}}$ &  & $10^{-14}{\rm~erg~cm^{-2}~s^{-1}}$ & $10^{44}{\rm~erg~s^{-1}}$ & Reduced $\chi^{2~}_{min}$  &$\nu$  \\ \hline
2 &    A  &  $9.60_{-7.00}^{+10.17}$ &  $2.02_{-0.44}^{+0.55}$ &   7.06 &   0.16 &  0.695 & 25
\\
  &    B &  7.9(fixed) & $1.93_{-0.15}^{+0.15}$ &   7.10 &  0.15 &  0.673 & 26
\\
3 &   A &  $3.72_{-3.72}^{+5.29}$ &  $2.28_{-0.27}^{+0.31}$ &   8.81 & 5.8 &  0.888 & 15 
\\
  &  B  &  7.9(fixed) &  $2.51_{-0.11}^{+0.12}$ &   8.85 & 7.56 &  0.944 & 16
\\
7 &  A  &  $20.66_{-9.69}^{+13.60}$ &  $3.00_{-0.48}^{+0.60}$ &   4.74 & 9.1 &  0.661 & 22
\\
  &  B  &  7.9(fixed) &  $2.38_{-0.14}^{+0.15}$ &   4.38 & 4.21 &  0.864 & 23
\\
  &  C & 7.9(fixed),$97.07_{-81.27}^{+120.13}$ & $3.06_{-0.61}^{+0.82}$ & 4.55   &18.8  & 0.742 & 22
\\
10 &   A  &  $18.1_{-4.7}^{+6.0}$ &  $3.6_{-0.3}^{+0.3}$ &   8.8  &  1.71 &  0.96 & 24
\\
   &   B  &  7.9(fixed) &  $3.1_{-0.1}^{+0.1}$ &   8.22 &   0.72 &  1.58 & 25
\\
  &  C  &  7.9(fixed), $28.0_{-8.0}^{+17.1}$ & $3.7_{-0.3}^{+0.3}$ & 8.8 & 1.9 & 0.96 & 24
\\
11 &  A &  $9.7_{-9.7}^{+16.1}$ &  $2.06_{-0.26}^{+0.66}$ &   4.10  &  1.54 &  1.201 & 15
\\
   &  B  &  7.9(fixed) &  $1.98_{-0.17}^{+0.16}$ &   4.06 &   1.43 &  1.130 & 16
\\
13 &  A  &  $17.69_{-13.01}^{+20.08}$ &  $2.70_{-0.64}^{+0.82}$ &   3.70 &   14.2 &  1.091 & 17
\\
   & B  &  7.9(fixed) &  $2.25_{-0.17}^{+0.19}$ &   3.51 &   8.45 &  1.105 & 18
\\
15 & A  &  $18.72_{-7.39}^{+9.83}$ &  $3.18_{-0.39}^{+0.47}$ &   6.32 &   13.2 &  1.112 & 15
\\
   & B  &  7.9(fixed) &  $2.62_{-0.12}^{+0.13}$ &   5.93 &   6.35 &  1.458 & 16 \\
   & C  &  7.9(fixed),  $95.3_{-74.2}^{+100.4}$ & $3.3_{-0.5}^{+0.7}$ & 6.14 & 36.10 & 1.225 & 15 
\\
17 &  A  &  $9.85_{-9.85}^{+20.05}$ &  $1.98_{-0.66}^{+0.83}$ &   2.43 &  5.80 &  1.010 & 8
\\
   &  B  &  7.9(fixed) &  $1.89_{-0.27}^{+0.25}$ &   2.41 &  5.39 &  0.903 & 9
\\
18 & A  & $16.62_{-8.74}^{+25.07}$ &  $2.28_{-0.85}^{+1.09}$ &   2.36 &   6.91 &  1.250 & 15
\\
   & B  &  7.9(fixed) &  $1.92_{-0.34}^{+0.29}$ &   2.26 &   4.93 &  1.205 & 15
\\
20 & A  &  $17.41_{-13.19}^{+22.40}$ &  $2.79_{-0.68}^{+0.98}$ &   2.76   &  7.4 &  0.825 & 13
\\
   &B   &  7.9(fixed) &  $2.35_{-0.20}^{+0.23}$ &   2.62   &  4.31 &  0.850 & 14
\\
21 &A   &  $10.86_{-10.86}^{+44.15}$ &  $1.68_{-1.07}^{+1.57}$ &   1.56 & 2.0 &  0.736 & 5
\\
   & B  &  7.9(fixed) &  $1.55_{-0.65}^{+0.42}$ &   1.55 & 1.85 &  0.618 & 6
\\
23 & A  &  $5.28_{-5.28}^{+10.00}$ &  $2.68_{-0.44}^{0.56}$ &   4.25 & 2.68 &  1.055 & 12
\\
   & B  &  7.9(fixed) &  $2.83_{-0.20}^{+0.25}$ &   4.24 & 3.24 &  0.995 & 13
\\
24 & A   &  $1.56_{-1.56}^{+22.38}$ &  $1.69_{-0.38}^{+1.00}$ &   1.96 &   3.51 &  0.663 & 8
\\
   & B  &  7.9(fixed) &  $2.00_{-0.33}^{+0.31}$ &   1.96 &   4.46 &  0.731 & 9
\\
29 & A  &  $4.22_{-4.22}^{+18.46}$ &  $2.49_{-0.48}^{+0.93}$ &   3.07 &  11.35 &  0.545 & 12
\\
   & B  &  7.9(fixed) &  $2.70_{-0.26}^{+0.33}$ &   3.06 &  14.5 &  0.519 & 13
\\
30 & A  &  $<13.75$ &  $1.44_{-0.34}^{+0.65}$ &   1.86 &   4.81 &  1.350 & 8
\\
   &  B  &  7.9(fixed) &  $1.79_{-0.39}^{+0.33}$ &   1.79 &   5.90 &  1.351 & 9 \\
31 &  A  &  $<855.7$ &  $1.95_{-1.22}^{+4.12}$ &   1.47 &   9.35 &  0.830 & 7
\\
   &  B  &  7.9(fixed) &  $1.42_{-0.56}^{+0.39}$ &   1.41 &   6.17 &  0.778 & 8
\\
37 &  A &  $<35.54$ &  $2.87_{-0.89}^{+1.59}$ &   1.42 &  4.29 &  1.167 & 10
\\
   & B  &  7.9(fixed) &  $2.84_{-0.41}^{+0.69}$ &   1.42 &  4.15 &  1.061 & 11
\\
48 &A  &  $<5.02$ &  $1.96_{-0.23}^{+0.38}$ &   2.79 &  0.46 &  0.747 & 15
\\
   & B &  7.9(fixed) &  $2.43_{-0.28}^{+0.32}$ &   2.62 &  0.70 &  0.971 & 16
\\
56 & A &  $22.26_{-20.66}^{+48.62}$ &  $3.22_{-1.09}^{+1.92}$ &   1.73 &   17.4 &  1.343 & 13
\\
   & B  &  7.9(fixed) &  $2.53_{-0.30}^{+0.29}$ &   1.57 &   6.79 &  1.319 & 14
\\
57 &  A  &  $<10.58$ &  $2.36 _{-0.39}^{+1.00}$ &   1.91 &   2.77 &  0.652 & 15
\\
   & B    &  7.9(fixed) &  $2.96_{-0.44}^{+0.88}$ &   1.74 &  4.93  &  0.721 & 16
\\
61 & A &  $<9.73$ &  $1.89_{-0.31}^{+0.64}$ &   2.55 &  42.3 &  0.854 & 7
\\
   & B &  7.9(fixed) &  $2.37_{-0.35}^{+0.41}$ &   2.44 &  63.4 &  1.021 & 8
\\
63 & A &  $<5.96$ &  $1.96_{-0.31}^{+0.46}$ &   1.53 &    10.4 &  0.933 & 6
\\
   & B  &  7.9(fixed) &  $2.40_{-0.36}^{+0.45}$ &   1.45 &    15.6 &  1.296 & 9
\\
\hline
\end{tabular}
$~^a$Model A(B) a is redshifted simple power-law model modified by an intervening medium (Galaxy) described by \NH (Galactic \NH)  at z=0.0. Model C is same as model B with an addition absorbing component at the source redshift. \\
$~^b$Observed flux in units of $10^{-14}{\rm ~erg~cm^{-2}~s^{-1}}$ in the energy band $0.1-2.0{\rm ~keV}$. \\
$~^c$Unabsorbed soft X-ray luminosity in units of $10^{44}{\rm ~erg~s^{-1}}$ in the energy band $0.1-2.0{\rm ~keV}$. \\
\end{flushleft}
\end{table*}

{\begin{figure*}
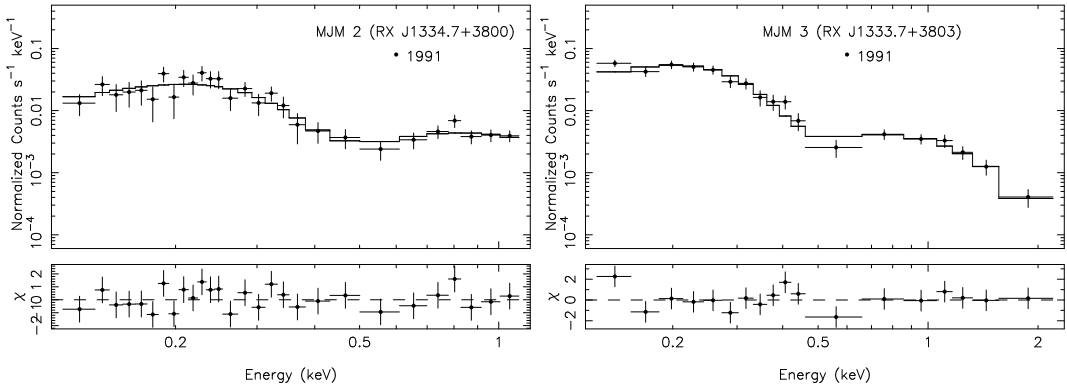

	\centering
	\includegraphics[angle=0,width=7cm]{m2_xspec.ps}
	\includegraphics[angle=0,width=7cm]{m3_xspec.ps}
	\caption{Examples of the $ROSAT$ PSPC spectra of MJM QSOs observed on 1991 and the best-fit models. 
The fitted model to each QSO spectrum is a redshifted 
power-law absorbed by our own Galaxy. } 
	\label{xray_spectra}
	\end{figure*}}

\begin{table*}
\caption{F-test for the presence of excess absorption at $z=z_{QSO}$ over the Galactic absorption of N$_H=7.9x10^{19}cm^{-2}$ at $z=0$ derived for the 1991 observation}
\label{ftest}
\begin{flushleft}
\begin{tabular}{cccc}
\hline
 MJM No. & F-statistics  value & F-statistic probability & dof\\
\hline
 7       &  4.79  & $3.95\times10^{-2}$ & 22\\
 10      & 18.0   & $2.83\times10^{-4}$ & 24 \\
 15      & 4.04   & $6.28\times10^{-2}$ & 15 \\
\hline   
\end{tabular}
\end{flushleft}
\end{table*}

For the QSOs MJM~29, MJM~30, MJM~48, MJM~61, and MJM~63, the amount of the 
absorbing column derived from the power-law fitting is unphysically small compared 
to the Galactic values of absorbing columns along their respective lines of sight. 
In order to constrain the photon indices, we fixed the absorbing column density 
to the Galactic value along the respective sight lines of individual QSOs and 
carried out the power-law model fitting to the PSPC spectra of the above QSOs. 
The PSPC spectra of the QSOs are well described by a 
power-law modified with absorbing column in our own Galaxy, as evidenced by the 
reduced $\chi^2_{min}$ values (see Table~\ref{fit91_result}).  

\begin{table}
\caption{Spectral model parameters of QSOs derived from the observation of 1993}
\label{fit93_result}
\begin{flushleft}
\begin{tabular}{cccccccc}
\hline
MJM No. & Model$^a$ & ${\rm N_{H}}$ & $\Gamma$ & $f_{obs}^b$ & $L_{unabs}^c$  & Minimum  & dof \\
        &       & $10^{19}{\rm~cm^{-2}}$ &  &  $10^{-14}{\rm~erg~cm^{-2}~s^{-1}}$ & $10^{44}{\rm~erg~s^{-1}}$ & Reduced $\chi^{2~}_{min}$  &$\nu$  \\ \hline
2 &  B & 7.9 & $1.82_{-0.11}^{+0.10}$ & 14.27   &  0.3 &0.751 & 23 \\
3 &  B  &  7.9 &  $2.46_{-0.24}^{+0.27}$ &   6.47 &  5.5 &  1.007 & 10
\\
7 &  B  &  7.9 &  $2.64$ &   7.13 &  7.8 &  2.123 & 18
\\
  &  C & 7.9, $286.5_{-156.7}^{+247.9}$ & $4.74_{-1.11}^{+1.69}$ & 7.64 & 994.0  & 1.344 & 17 \\
10 & B &  7.9       & $3.2_{-0.2}^{+0.2}$ & $7.9$ & 1.1  & 1.11 & 12 \\
   & C   & 7.9, $2.2_{-1.8}^{+2.6}$ & $3.7_{-0.5}^{+0.6}$ & $8.2$ &2.9  & 0.82 & 11 \\
11   & B  &  7.9 &  $2.23_{-0.27}^{+0.28}$ &   3.85 &  1.5 &  0.802 & 10
\\
13 &  B & 7.9 & $1.64_{-0.63}^{+0.3}$  & 2.08 & 4.14 & 1.330 & 5
\\
15   & B &  7.9 &  $2.96_{-0.32}^{+0.44}$ &   2.35 &  11.8 &  0.760 & 10
\\
17  & B  &  7.9 &  $2.11_{-0.24}^{+0.24}$ &   3.53 & 8.6 &  0.664 & 6
\\
18   & B &  7.9 &  $1.57_{-0.62}^{+0.42}$ &   3.70 &  7.4 &  0.768 & 7
\\
20   &  B &  7.9 &  $1.84_{-0.57}^{+0.51}$ &   2.27 &  3.1 &  1.520 & 5
\\
21   & B  &  7.9 &  $1.48_{-0.77}^{+0.47}$ &   2.67 &  3.2 & 1.245  & 9
\\
23  &  B &  7.9 &  $2.7_{-0.28}^{+0.37}$ &   3.84 &   2.8 &  0.454 & 6
\\
24   & B &  7.9 &  $2.06_{-0.41}^{+0.43}$ &   2.96 &  6.9 &  1.117 & 4
\\
29   & B  &  7.9 &  $2.55_{-0.25}^{+0.30}$ &   4.11 &   18.5 &  1.486 & 7
\\
30   &  B &  7.9 &  $2.05_{-0.36}^{+0.71}$ &   2.55 &   9.2 &  0.629 & 5
\\
48   &  B &  7.9 &  $1.73_{-0.79}^{+0.49}$ &   2.44 &  0.5 &  0.669 & 6
\\
57   & B  &  7.9 &  $2.35_{-0.42}^{+0.44}$ &   2.29 &   4.9 &  1.335 & 5
\\
\hline
\end{tabular}
$~^a$Models B and C are same as that defined in Table~\ref{fit91_result}. \\
$~^b$Observed flux in units of $10^{-14}{\rm ~erg~cm^{-2}~s^{-1}}$ in the energy band $0.1-2.0{\rm ~keV}$. \\
$~^c$Unabsorbed soft X-ray luminosity in units of $10^{44}{\rm ~erg~s^{-1}}$ in the energy band $0.1-2.0{\rm ~keV}$. \\
\end{flushleft}
\end{table}

In order to investigate any change in the soft X-ray spectral shape and flux, 
we  have also analyzed the PSPC spectra of the QSOs observed in 1993. 
The PSPC spectra of the QSOs and the corresponding background spectra were 
extracted from the observation of 1993 using the similar extraction regions as for the 
observation of 1991. The same spectral models, namely redshifted and absorbed power-law 
model with \NH as a free parameter and \NH fixed to the Galactic, as described above 
were fitted to the PSPC spectra obtained in 1993. The power-law model with \NH as a 
free parameter is a good description to the observed PSPC spectra of individual QSOs. 
Except for the QSOs MJM~7 and MJM~10, the Galactic \NH along the line of sight of each 
QSO was found to be well within the range of derived \NH. Therefore, the power-law 
model with \NH fixed to the Galactic value was fitted to the spectrum of each QSO. 
The best-fit parameters derived are given in Table~\ref{fit93_result}.

The PSPC spectrum of the QSO MJM~7 is well described by a power-law of 
$\Gamma= 3.91_{-0.58}^{+1.38}$, and $\NH= 3.24_{-1.19}^{+1.70}\times10^{20}{\rm~cm^{-2}}$ 
(reduced $\chi^2_{min} =1.262$ for 17 dof). The best-fit value of \NH is much higher 
than the Galactic value ($N_H^{Gal} = 7.9\times10^{19}{\rm~cm^{-2}}$)  
suggesting excess absorption. The power-law fit with \NH fixed to the Galactic 
value is not acceptable (reduced $\chi^2_{min} = 2.123$ for 18  dof). 
Addition of an absorbing component, apart from the Galactic absorption 
component, local to the QSO gives an acceptable fit  
(reduced $\chi^2_{min} = 1.344$ for 17 dof) 
 The best-fit parameters are $\Gamma_X = 4.74_{-1.11}^{+1.69}$, and 
 $\Delta \NH = 2.86_{-1.57}^{+2.48}\times10^{21}{\rm~cm^{-2}}$. The F-statistic 
 value for the addition of the intrinsic absorption component to the power-law with 
 \NH fixed at the Galactic value is 11.4 and the F-statistic probability is 
 $3.55\times10^{-3}$. Thus the addition of an excess absorption at the QSO redshift to 
 the power-law model with fixed \NH is a significant improvement at $99.4\%$ level. 

The power-law model with \NH as a free parameter is an acceptable fit 
(reduced $\chi^2_{min} = 0.83$ for 11 dof) to the PSPC spectrum of MJM~10. 
The best-fit parameters are $\NH = 16.0_{-6.4 }^{+9.1}\times10^{19}{\rm~cm^{-2}}$ 
and $\Gamma_X = 3.7_{-0.4}^{+0.5}$. The best-fit value of \NH is significantly higher 
than the Galactic value ($N_H^{Gal}=7.9\times10^{19}{\rm~cm^{-2}}$) indicating 
excess absorption. The best-fit power-law model with \NH fixed to the Galactic 
value is also an acceptable fit (reduced $\chi^2_{min} = 1.11$ for 12 dof) 
although the fit is becomes poor again indicating excess absorption. Next, 
we introduced an additional absorbing component local to the source and 
carried out the model fitting. The fit is an acceptable fit 
(reduced $\chi^2_{min} = 0.82$ for 11 dof).  
The F-statistic value for the addition of new component to the power-law model 
with fixed \NH is 5.24 and the corresponding F-statistic probability is 
$4.28\times10^{-2}$. Hence the power-law model modified by Galactic absorption and 
intrinsic absorption is a  significant improvement at $90\%$ level but not a 
significant improvement at $95\%$ level over the power-law model with Galactic 
absorption. In the case of MJM~15, an excess absorption is not 
inferred from the 1993 observations. This discrepancy could be due to the low S/N and poor spectral resolution of the PSPC spectra of MJM~15.

\subsection{Optical Spectroscopy}
Optical spectra of 12 of the 22 QSOs were obtained on the night of
1994 April 7 with the Multi Object Spectrograph (MOS) on the 3.6-m
Canada France Hawaii Telescope (CFHT) as a part of a program to
identify X-ray sources (McHardy \etal 1998). A ${\rm 300~l~mm^{-1}}$
grism  in the first order with a Lorel3 CCD detector was used to cover
a wavelength range of $4000{\rm~\AA}$--$9000{\rm~\AA}$ with
$\sim15{\rm~\AA}$ resolution.

Optical spectra of the 12 QSOs were reduced using IRAF (v2.11.3). For details of the reduction see McHardy et al (1998). The flux
calibration is somewhat uncertain because the
slitlets were not aligned at the parallactic angle. The flux calibrated 
spectra were corrected for Galactic extinction by adopting the extinction law 
in Cardelli \etal (1989). Color excess $E_{B-V}$ due to Galactic reddening along the line of sight
of each QSO was calculated from the neutral hydrogen column density $\NH$ using the relation
\begin{equation}
\NH=4.93\times 10^{21} \times E_{B-V}{\rm ~cm^{-2}}
\end{equation}
(Diplas \& Savage 1994). The dereddened spectra were then 
deredshifted by dividing the starting wavelength and the pixel
width in Angstrom by (1+z). The final spectra of 12 QSOs are shown
in Figure~\ref{optical_spectra}. For each of the 12 objects, relative flux against the
rest wavelength in ${\rm \AA}$ has been plotted. Strong emission 
lines have been marked in the figure. 

The analysis of the emission lines of MJM~10 has been discussed by
Dewangan \etal (2001). Emission lines of other QSOs were modeled by Gaussian profiles using the IRAF task {\it ngaussfit}. As a first step, 
we fitted a single Gaussian profile to the strong emission lines. Most of the emission lines are described by a single Gaussian profiles, however, the broad wings of the Mg~II lines in the spectra of the QSOs MJM~3, 7, and 20 are not well fitted by a single Gaussian profile. Therefore, the Mg~II line in the spectra of the above QSOs were modeled by two Gaussian which gave satisfactory fit to the line core as well as to the wings. The two component fitting to the Mg~II line revealed that 
the broad component is blue shifted with respect to the narrow component (see Table~\ref{fit_optical}). The broad components of the Mg~II line is blue shifted by $\sim1180$, $\sim790$, and $\sim1080{\rm~km~s^{-1}}$ in the spectra of MJM~3, 7, and 20, respectively. In the rest of the QSO spectra only one component is detected. This could be due to the low S/N ratios of the optical spectra.

{\begin{figure*}
	\centering
	\includegraphics[angle=0,width=15cm]{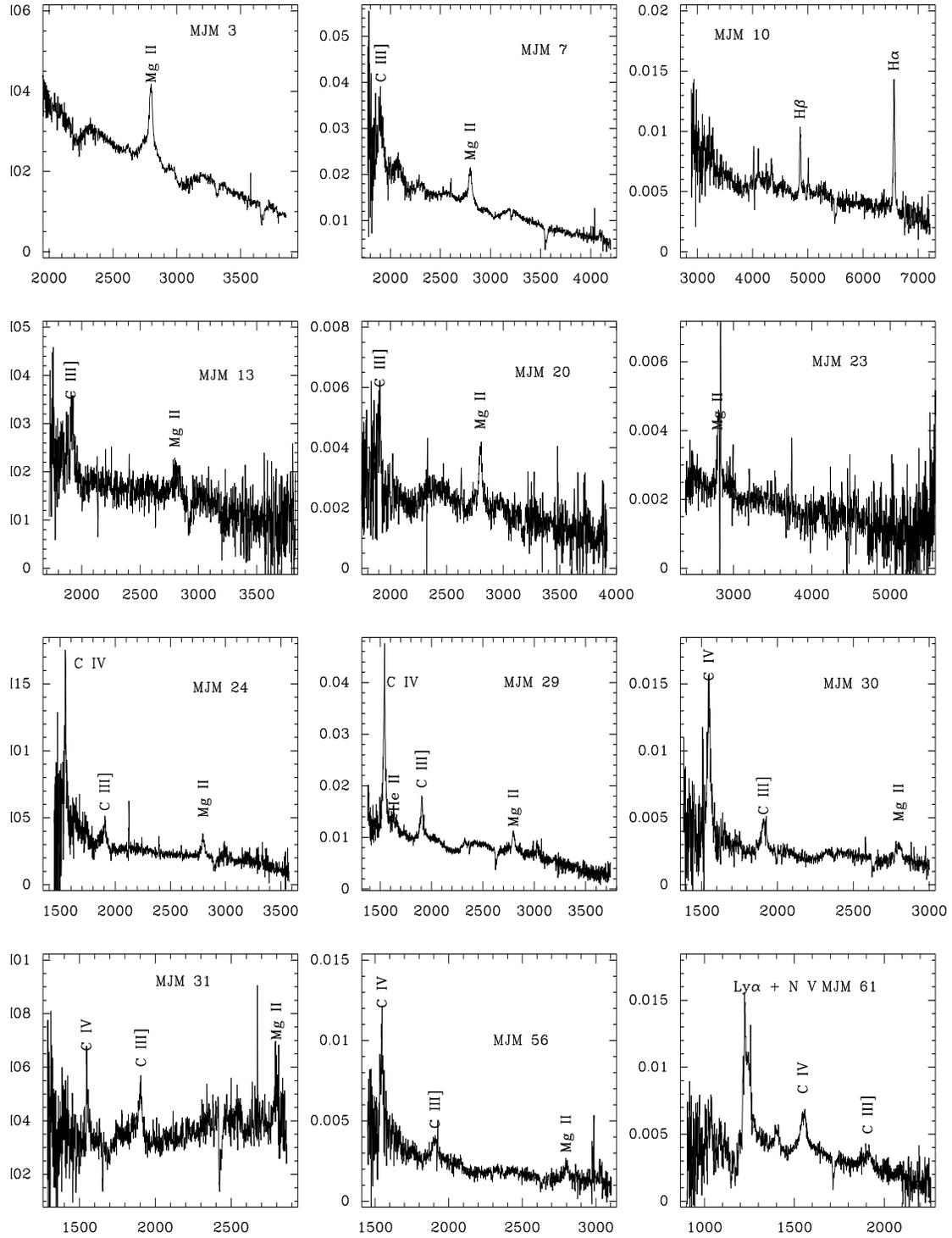}
	\caption{Optical spectra of 12 of the 22 MJM QSOs. Vertical scale is the relative flux and the horizontal axis is the rest wavelength. The absorption feature seen in all the spectra is due to atmospheric absorption.}
	\label{optical_spectra}
	\end{figure*}}

\begin{table*}
\caption{Emission line parameters of the QSOs. The quantities in brackets represent $1\sigma$ errors.}
\label{fit_optical}
\begin{flushleft}
\begin{tabular}{ccccccccc}
\hline
MJM & Line &  \multicolumn{3}{c}{Component 1} &  \multicolumn{3}{c}{Component 2} & Comments \\
 No.     &       & $\lambda_c$ & $\Delta v_{FWHM}^1$ & EW & $\lambda_c$ & $\Delta v_{FWHM}^1$ & EW &  \\
         &       & (${\rm~\AA}$)     & (${\rm~km~s^{-1}}$) & (${\rm~\AA}$) & (${\rm~\AA}$) & (${\rm~km~s^{-1}}$) & (${\rm~\AA}$) & \\
\hline
3  & Mg~II & $2798.4$ & $2567.8$ & $73.3$ & $2786.7$ & $11684.7$ & $160.0$ &  --\\
   &       & (0.2)   & (56.9) & (3.2) & (0.8) & (346.3) & (7.1) & -- \\
\\
7 & Mg~II & $2798.5$ & $2834.1$ & $53.7$ & $2791.1$ & $9788.6$ & $93.4$ & --\\
  &       & (0.6) & (144.8) & (4.1) & (2.0) & (722.9) & (7.9) & --\\ 
  & C~III] & $1895.5$ & $12048.1$ & $221.7$ & -- & -- & -- & Blended with Si~III]$\lambda1892$ \\ 
  &        &          &           &         &    &    &    &     and Al~III$\lambda1857$ \\
  &        & (3.2) & (1558.1) & (33.8) & -- & -- & --&  -- \\
\\
10 & H$\beta$  & $4861$   &$2853$  & $15.8$ & $4861$ &  $877$ & $19.5$ & from Dewangan \etal (2001) \\
   &      & -- &  (520) & (3.6) & -- & (91) & (3.8) & \\
   & ${\rm [O~III]\lambda5007}$ & 5007 & -- & -- &-- &  -- & $5.0$  & ''   \\
   &                    & -- & -- & -- & -- & -- & (0.5) & '' \\
   &  H$\alpha$ & 6563  & $2843$ & $33.8$ & 6563 &  $883$ & 54.5 &  ''  \\
   &           & -- & (479) & (7.5) & -- & (41) & -- &  '' \\
   & Fe~II     & -- & -- & 159 & -- & --  &-- & '' \\
\\
13 & --   & --        &  --    & -- & -- & -- & --&  Line parameters unreliable \\
\\
20 & Mg~II & $2798.0$ & $2694.4$ & $117.5$ & $2787.9$ & $9969.5$ & $147.5$ & C~III] parameters unreliable \\
   &        & (0.7) & (244.9) & (11.3) & (3.4) & (1264.0) & (19.8) &  \\
\\
23 &  --   &  --       &  --      & --  & -- & --&-- & Line parameters unreliable \\
\\
24 & Mg~II & 2795.9 & $3340.9$ & $116.9$ & -- & -- & --& Broad component undetectable \\
   &       & (0.6)       & (123.0) & (2.7)  & -- & -- & -- & --  \\ 
& C~III] & -- & -- & $51.7$ &-- & --  & -- & FHHMs unreliable.\\
  &      & -- & -- & (2.3) & -- & -- & -- & \\
\\
29 & Mg~II & $2797.0$ & $3363.5$ & $109.1$ & -- &-- &-- & Broad component not detectable \\
   &       & (0.3) & (69.2) & (4.8) &-- &-- &--  &    C~III] profile corrupted\\  
\\
30 & Mg~II & 2798.6 & $5273.7$ & $233.4$ & -- & -- & -- & Only one component detected\\
   &       & (0.8)  & (318.9) & (124.2) & -- & -- & -- &  C~III] line corrupted\\    
   & C~IV  & $1548.5$ & $4996.9$ & $544.8$ & -- & -- & -- & --  \\
   &       & (0.2) & (101.3) & (106.4) &--&-- &--&--\\
\\
31 & Mg~II & $2797.5$ & $2905.3$ & $83.9$ &-- &-- &-- &-- \\
   &       & (1.1) & (231.6) & (32.2) &--&--&--& -- \\
   & C~III] & $1901.5$ & $4945.8$ & $116.4$ &-- &--&--& Deblending not possible \\
   &       &  (0.6) & (388.0) & (16.2) &--&--&-- & -- \\
\\
56 & Mg~II & $2797.7$ & $4408.3$ & $184.7$ &-- &-- &-- & C~III] corrupted   \\
   &       & (1.3)         & (330.4) & (40.7) &-- &-- &-- &-- \\ 
\\   
61 & C~III] & $1907.9$ & $11603.9$ & $377.4$ &-- &-- &-- &-- \\
   &        & (3.2)   &  (1106.8) & (269.7) &--  &-- & -- &--  \\       
   & C~IV   & $1550$ & $8282.1$ & $318.5$ &-- &-- &-- & Ly$\alpha$ + N~V is corrupted \\
   &        & (0.4) & (172.6) & (53.7) &-- &-- &-- & -- \\
\\
\hline
\end{tabular}
\newline
$~^1$Corrected for the instrumental broadening (FWHM$\sim15{\rm~\AA}$).
\end{flushleft}
\end{table*}

\section{Discussion}

{\begin{figure*}
	\centering
	\includegraphics[angle=-90,width=12cm]{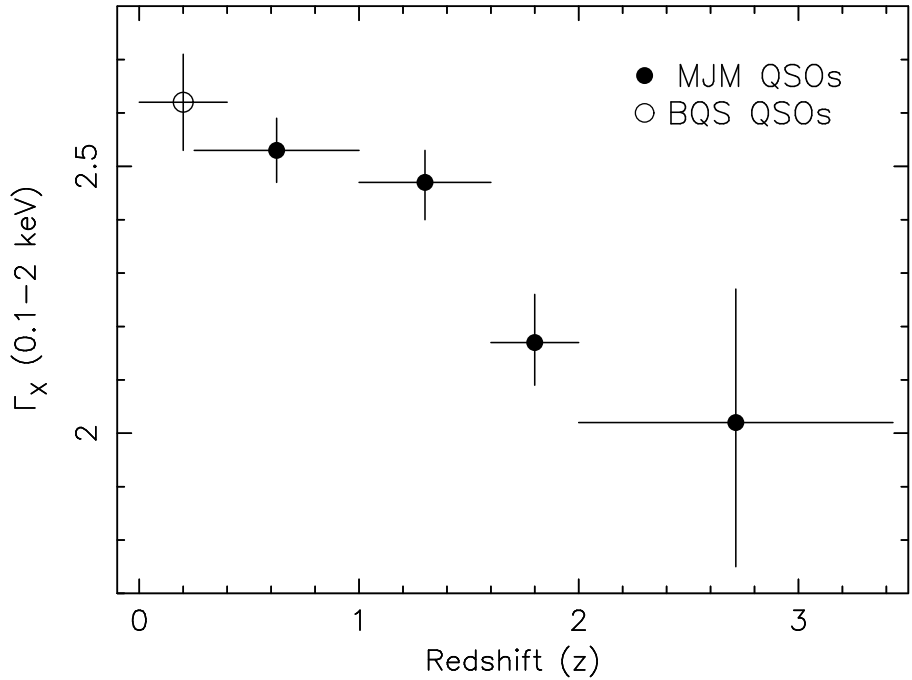}
\caption{Best-fitting power-law photon indices obtained by fitting simple power-law models to the average spectra of MJM QSOs at four redshift bins. Also plotted is the average photon index of 23 QSOs from the BQS sample of Laor \etal(1997). The average photon index of QSOs flattens at higher redshift.}
\label{z_pi}
\end{figure*}}

\subsection{The soft X-ray spectral shape}
The deep $ROSAT$ PSPC observations together with improved sensitivity and energy resolution,
compared with earlier instruments, allows us to determine the shape of the soft X-ray continua
of QSOs of redshifts up to 3.43.  
There are three QSO namely MJM~7, MJM~10, and MJM~15 which show steeper soft X-ray 
continua ($\Gamma_{X} > 3.0$). The QSO MJM~10 is known to be a narrow-line 
QSO (NLQSO) -- high luminosity version of narrow-line Seyfert~1 
galaxies (NLS1s) (Dewangan \etal 2001). 
The average photon index of the QSOs 
is $2.40\pm0.09$ with a dispersion of $0.57$. 10 QSOs with $z>1.6$ have mean photon index of $2.12\pm0.10$ (with dispersion about mean of $0.39$) and 12 QSO with $z<1.6$ have mean photon index of $2.62\pm0.15$ with a dispersion of $0.61$.
The average photon index of the QSOs with $z>1.6$ in our sample appears to be flatter
than the average photon index of QSOs at low redshifts. 
Walter \& Fink (1993)
analyzed the PSPC spectra of 58 AGNs observed in the $ROSAT$ All Sky
Survey (RASS) which contains a subsample of 24 QSOs with mean photon
index, $\Gamma_X = 2.57\pm0.06$ (Laor \etal 1997). 
A similar average photon index of $2.65\pm0.07$ was found by Schartel \etal (1996) for 72 QSOs from the LBQS sample detected
in the RASS. 
Laor \etal (1997)
studied the $ROSAT$ PSPC spectra of a complete sample of 23
optically selected QSOs
with $z\le0.4$ and found average photon index of $2.62\pm0.09$ for
their sample.

In order to further explore the redshift dependence of soft X-ray spectral slope of QSOs, 
we formed average spectra of QSOs in four redshift bins -- $z=0.25-1.0$ (5 QSOs), $z=1.0-1.6$ (7 QSOs), $z=1.6-2.0$ (7 QSOs), and $z=2.0-3.43$ (3 QSOs). The average spectra were then fitted by simple 
power-law modified by the Galactic absorption only.  
In Figure~\ref{z_pi}, we 
have plotted the average photon index as a function of redshift for the QSOs in our sample. Also plotted in Fig.~\ref{z_pi} is the average photon index of 23 Quasars from the Bright Quasar Survey (BQS) sample of Laor \etal (1997). The average photon index flattens at higher redshifts e.g., $<\Gamma_X> = 2.62\pm0.09$ at $z \le 0.4$ and $<\Gamma_X> = 2.17_{-0.08}^{+0.09}$ at $1.6 \le z \le 2.0$. 

Blair \etal (2000) studied soft X-ray spectral evolution with redshift using 
a large sample of 165 QSOs observed with {\it ROSAT} and found the $0.1-2{\rm~keV}$ 
average spectra in 5 redshift bins harden from $\Gamma_X \sim 2.6$ at $z=0.4$ to 
$\Gamma_X \sim 2.1$ at $z=2.4$. This result is similar to that found here. 
Blair \etal (2000) also noted that the spectra in $0.5-2{\rm~keV}$ band show 
no significant variation in spectral index with redshift suggesting the presence 
of a spectral upturn below $0.5{\rm~keV}$. While Blair \etal (2000) found that at lower redshifts spectra need a soft excess component in addition to a power-law and inclusion of a blackbody component ($kT \sim 100{\rm~eV}$) for the soft excess results in a significant improvement over a single power-law, Laor \etal (1997) did not require such a soft excess component. The discrepancy could be due to the fact that Laor \etal (1997) added $1\%$ systematic errors in quadrature to the statistical errors.  The difference between the average spectral slopes of QSOs at low and high redshifts  can be understood in terms 
of mean X-ray spectrum consisting of a power-law of $\Gamma_X \sim 1.9$ and a soft 
excess below $\sim 0.5{\rm~keV}$ in the rest frame. At higher redshifts ($z\la2$) the 
rest frame soft excess component below $\sim 0.5{\rm~keV}$ is not covered in the 
observed $0.1-2{\rm~keV}$ band, the observed spectrum is a power-law with 
$\Gamma_X \sim 1.9$. While at lower redshifts, the observed spectrum, consisting 
of both soft excess and power-law component, is steeper when fitted with a single 
power-law and similar to that of $z \la 2$ QSOs when the soft excess component 
below $\sim 0.5{\rm~keV}$ is excluded from the fit. This could also be the reason why 
Reeves \& Turner (2000) found mean photon index of $\sim 1.9$ in the 
$0.6-10{\rm~keV}$ {\it ASCA} band for their sample of 27 radio quiet quasars.

\subsection{Intrinsic Absorption}
 Due to the low signal-to-noise of the PSPC spectra of the QSOs in our 
 sample, we are not able to determine the consistency of HI 
 columns derived from the 21~cm radio observations 
 (Dickey \& Lockman 1990) and HI columns derived from the 
 absorbed power-law  model fits to the observed PSPC spectra. 
 Laor \etal (1997) have shown that the two columns agree to a
 level of about $5\%-9\%$ at high Galactic latitudes. There are indications that some broad-line AGNs show intrinsic absorption (e.g., Page \etal 2001; Akiyama \etal 2000). The power-law
 models modified by the Galactic absorption provide acceptable fits
 to the PSPC spectra of all QSOs except MJM~7, MJM~10, and MJM~15 
 (see Tables~\ref{fit91_result},\&~\ref{ftest}). Thus most of the QSOs in our sample lack 
 intrinsic absorption. Only three QSOs, MJM~7, MJM~10, and MJM~15, out of
 22 show indications for the presence of intrinsic 
 absorption. The amount of intrinsic absorption inferred from the
 best-fit absorbed power-law models, $\sim 10^{21}{\rm~cm^{-2}}$ for MJM~7, is quite 
 high and may cause the QSO to be redder than 
 others in the sample. The ratio of soft X-ray to optical R band flux can be
 written as
 \begin{equation}
 log(\frac{f_{X}}{f_{R}})=log(f_{X})+\frac{m_{R}}{2.5} + 3.5
 \end{equation}
 where $f_{X}$ and $f_{R}$ are X-ray and optical R band fluxes, and $m_{R}$ is the R band magnitude. For MJM~7, the ratio of observed X-ray and optical flux ($\frac{f_{X}}{f_{R}}$) is calculated to be $2.95\times10^{-3}$ while the ratio is 0.0160 for MJM~15 which is located at the same redshift (z=1.14) as MJM~7. In order to investigate whether the lack of soft X-ray emission of MJM~7 is only due to intrinsic absorption or the QSO is intrinsically X-ray weak, we calculated the ratio of unabsorbed soft X-ray flux and observed R band flux for MJM~7 and MJM~15. The ratio ($\frac{f_{X}}{f_{R}}$)  is 0.017 for MJM~7 and 0.072 for MJM~15. Thus, the low observed X-ray flux of MJM~7 may not be entirely due to absorption and MJM~7 could be intrinsically weaker. 
 The amount
 of intrinsic HI column in MJM~10 is small ($2.8_{-0.8}^{+1.7}\times10^{20}{\rm~cm^{-2}}$) and
 is similar to that observed in some NLS1 galaxies (Grupe \etal 1998). 

\subsection{Optical Spectra}
Optical spectra of the 12 QSOs, except MJM~10, are typical of QSOs. 
The optical spectra of MJM~10 is typical of NLS1 galaxies or NLQSOs and 
has been discussed in detail by Dewangan \etal (2001). The other 11 QSOs 
show broad permitted lines in the ultra-violet region. The Mg~II lines in the 
spectra of the QSOs MJM~3, 7, and 20 have been decomposed into a narrow and a 
broad component. The broad component is found to be blueshifted by $\sim1180$, 
$\sim 790$, and $\sim1080{\rm~km~s^{-1}}$ with respect to the narrow component 
in the spectra of MJM~3, 7, and 20.  
The Mg~II lines in the spectra of the QSOs MJM~3, 7, and 20 have
been decomposed into a narrow and a broad component. The broad
component is found to be blueshifted by $\sim1180$, $\sim 790$, and
$\sim1080{\rm~km~s^{-1}}$ with respect to the narrow component in the
spectra of MJM~3, 7, and 20. This suggests that  either the red wings of
the Mg~II lines of MJM~3, 7, and 20 are  affected by absorption features
more strongly than  the blue wings or the broad components of the Mg~II lines
originate in the outflowing winds.

\section{Conclusions}
We have derived soft X-ray spectral shapes and light curves of a nearly complete 
sample of 22 QSOs. We also presented optical spectra of 12 QSOs
from our sample. Our main results are as follows. \\
(i) About $33\%$ of the QSOs show a long term ($\sim 2{\rm~yr}$) 
soft X-ray variability while only one QSO MJM~10 shows rapid variability. \\
(ii) Only three QSOs, MJM~7, 10, and 15 out of 22 QSOs ($\sim7\%$) show 
indications  for the presence of significant intrinsic absorption. 
The former two QSOs show excess 
absorption during both the observations of 1991 and 1993, while MJM~15 does not 
show excess absorption during the observation 1993.  \\
(iii) The soft X-ray photon index of the QSOs in our  
sample ranges from 1.4 to 3.7. The average photon index of the sample 
is $2.40\pm0.09$ with a dispersion of $0.57$. \\
(iv) The average photon index of the QSOs is found to flatten at higher redshift. 
This can be 
understood in terms of the redshift effect of mean intrinsic QSO spectra
consisting two components -- a soft excess component and a power-law. \\
(v) Only one QSO MJM~10 out of 22 has been found to be an NLQSO. \\
(vi) The broad component of the Mg~II line in the spectra of MJM~3, 7, 
and 20 are found to be blueshifted by $\sim1180$, $\sim790$, and
$\sim 1080{\rm~km~s^{-1}}$, respectively.

\end{document}